\documentclass{article}

\usepackage[english]{babel}
\usepackage{xcolor}
\usepackage[letterpaper,top=2cm,bottom=2cm,left=3cm,right=3cm,marginparwidth=1.75cm]{geometry}
\usepackage{verbatim}
\usepackage{amsmath}
\usepackage{graphicx}
\usepackage[colorlinks=true, allcolors=blue]{hyperref}
\usepackage{subcaption}
\usepackage{cancel}
\usepackage{authblk}
\usepackage{csquotes}
\usepackage{xcolor}
\usepackage{ulem}

\usepackage{pdfpages}
\usepackage{authblk}
\usepackage[sorting = none, style=numeric-comp]{biblatex}

\DeclareFieldFormat{title}{#1}
\DeclareFieldFormat{author}{#1}
\DeclareFieldFormat{journaltitle}{#1}
\DeclareFieldFormat{date}{#1}

\DeclareFieldFormat{booktitle}{#1}

\DeclareFieldFormat{url}{Available from\addcolon\space\url{#1}}

\DeclareFieldFormat{note}{#1}

\title{Analyzing User Ideologies and Shared News During the 2019 Argentinian Elections}

\author{Sofía M del Pozo$^{1,}$$^2$, Sebastián Pinto$^{1,}$$^2$, Matteo Serafino$^3$,  Lucio Garcia $^{1,}$$^2$, Hernán A Makse$^3$ and Pablo Balenzuela$^{1,}$$^2$.

\par
{\normalsize $^1$ Universidad de Buenos Aires, Facultad de Ciencias Exactas y Naturales, Departamento de Física. Buenos Aires, Argentina.}
\par
{\normalsize $^2$ CONICET - Universidad de Buenos Aires, Instituto de Física Interdisciplinaria y Aplicada (INFINA). Buenos Aires, Argentina.}

\par
{\normalsize $^3$Levich Institute and Physics Department, City College of New York, 10031 New York, USA.
}
}

\begin{document}

\maketitle

\begin{abstract}

The extensive data generated on social media platforms allow us to gain insights over trending topics and public opinions. Additionally, it offers a window into user behavior, including their content engagement and news sharing habits. In this study,  we analyze the relationship between users' political ideologies and the news they share during Argentina's 2019 election period. Our findings reveal that users predominantly share news that aligns with their political beliefs, despite accessing media outlets with diverse political leanings. Moreover, we observe a consistent pattern of users sharing articles related to topics biased to their preferred candidates, highlighting a deeper level of political alignment in online discussions. We believe that this systematic analysis framework can be applied to similar scenarios in different countries, especially those marked by significant political polarization, akin to Argentina.

\end{abstract}

\section{Introduction}
In 1998 Richard Feynman wrote \textit{“The first principle is that you must not fool yourself, and you are the easiest person to fool.”} \cite{feynman1998cargo}. How we perceive things and subsequently respond to them is a phenomena potentially influenced by personal biases. 

The widespread use of social media platforms generates a large amount of data which, through careful interrogation and analysis, could reflect extensive and valuable information \cite{barbier2011data,newman2021reuters,digitalnewsreport2023}. This data not only sheds light on, for instance, trending topics \cite{chandrasekaran2020topics,lee2011twitter} and public opinions \cite{falkenberg2022growing,PaperSeba,anstead2015social,klavsnja2015measuring} but also provides insights into the individual characteristics of users based on their behavior, such as their interactions and the news they share \cite{tadesse2018personality,an2014sharing}. In particular, news sharing behavior on social media is a phenomenon worthy of study \cite{kalsnes2018understanding,kumpel2015news,lee2012news}, not only for its potential to infer users' information but also for its significant potential to influence society. Concerning the accuracy of shared news, the propagation of fake news could have serious implications, such as during elections \cite{allcott2017social,bovet2019influence} and the COVID-19 pandemic, where misinformation heightened anxiety and psychological distress \cite{rocha2021impact}. 

Numerous factors can influence the process of news sharing behavior \cite{lee2012news,kim2021people,karnowski2018users}. For example, Osmundsen et al. (2021) \cite{osmundsen2021partisan} demonstrated in their study that partisan polarization is the primary psychological motivation behind the sharing of political fake news on Twitter.
Westerwick et al. (2017) \cite{westerwick2017confirmation} examined the relationship between sources and content cues for confirmation bias, revealing that confirmation bias emerged irrespective of source quality \cite{westerwick2017confirmation}. In this sense, we can observe that user characteristics, in particular their political leaning or biases, can serve as both an explanation of news sharing behavior and as information resulting from this behavior. 

Bias is defined as the tendency to favour or dislike a person or thing, especially as a result of a preconceived opinion \cite{oxforddict}. While bias can manifest in different ways \cite{smith2014bias,delgado2004bias,kunda1990case,williams1975unbiased}, the two types of biases that specifically concern us in this study are those influencing news consumption behavior and those affecting the media on social networks, known as confirmation bias and media bias, respectively. As Raymond Nickerson explained in  \cite{nickerson1998confirmation}, confirmation bias refers to the inclination to search for or interpret evidence in a manner that aligns with pre-existing beliefs, expectations, or a currently held hypothesis. In essence, it represents an unintentional shaping of facts to fit one's hypotheses or beliefs. In the context of our research, confirmation bias can be recognized as an instance of the \textit{selective exposure theory}, as described by Stroud (2010) \cite{stroud2010polarization}, which elucidates individuals' propensity to prefer information that conforms to their pre-existing beliefs, while consciously avoiding contradictory content. Regarding news consumption research, media bias takes on a prominent role. Media  bias is defined as a deliberate and intentional tendency that favors a particular perspective, ideology, or desired outcome \cite{spinde2021automated,williams1975unbiased}.

As we  mentioned above, social media serves as a channel for news consumption, where several factors influence the dynamics of how these news are shared. In particular, both confirmation bias and media bias can interplay when the news that social media users read are shared by others who possess their own ideological biases. Therefore, the study of this dynamic is of significant importance due to the impact of news consumption on people's opinions \cite{mccombs1972, Guo2015, diaz2022echo,wei2011effective}, for example the consumption of biased news can influence voters’ decisions \cite{druckman2005impact}. Additionally, the interaction between social media, political polarization, and political disinformation can significantly shape a society's future, affecting the quality of public policy and its democratic principles \cite{tucker2018social}.

In this study, we examine the relationship between shared news and the ideologies of social media users who disseminate them during the 2019 general elections in Argentina. Specifically, we explore whether factors such as the news source, bias, or topics shared by users are associated with their political ideology. Our analysis incorporates data on the content of the shared news and the political affiliations of the users, previously categorized into Center-Left (CL) and Center-Right (CR) groups. The Twitter activity and partisan labels were obtained from the research conducted by Zhou et al. (2021) as referenced in \cite{zhou2021}.

The focal point of this study lies in examining the news shared by users within the existing dataset from \cite{zhou2021}. To collect this data, we performed web scraping of the text from the links of news articles shared by users. Following this, we evaluated the bias of these news articles, along with the bias of the news media and the topics they cover. Subsequently, we analyzed the correlation between these factors and the users' bias towards the candidates from the two primary coalitions competing for the presidency.

\section{Background}\label{Bck}

\subsection{Argentinian Context}
The systematic framework introduced in this work, aimed at quantifying both user and media outlet preferences, fills a significant gap in understanding, especially within the Argentine context. While some of the main Argentine media outlets are listed on Media Bias/Fact Check organization \cite{mediabiasfactcheck}, currently, there is no centralized source for evaluating the media bias of all outlets in the country. As we apply this framework to Argentina during the 2019 presidential election campaign, this section offers an overview of the political and media landscape during this period to provide contextualization.

\par Over the past decade, Argentina's political scene has been characterized by the predominance of two major coalitions: one, a center-Checkleft coalition (CL) led by Cristina Fernández de Kirchner, known as \textit{Frente de Todos}, and the other, a center-right coalition (CR) led by Mauricio Macri, referred to as \textit{Juntos por el Cambio}. Cristina Kirchner held the presidency in Argentina during the periods of $2007-2011$ and $2011-2015$, while Mauricio Macri served as president from 2015 to 2019, as documented in \cite{cantamutto2016}. During the 2019 elections, the center-left coalition presented Alberto Fernández and Cristina Fernández de Kirchner as their candidates. Meanwhile, the center-right coalition sought a second term for Mauricio Macri as president, with Miguel Ángel Pichetto as his vice-presidential candidate. National elections in Argentina comprise two obligatory phases: the primary election, known as PASO (which stands for \textit{Primarias, Abiertas, Simultáneas y Obligatorias} in Spanish, translating to Open, Simultaneous, and Obligatory Primaries in English), and the general election. In the year 2019, these events occurred on August 11th and October 27th, respectively. Additionally, if the results of the general election necessitate it, a third round, referred to as a \textit{ballotage}, may also be conducted.

\par Regarding the media landscape, the digital media scene in Argentina is primarily characterized by three major players: {\it Infobae}, {\it Clarín} and {\it La Nación}, each boasting approximately 20 million unique users in 2020, as reported by Comscore data \cite{comscore}. Following closely are a second tier of media outlets with audience numbers ranging from 6 to 13 million unique visitors. Prominent among this group are {\it Página 12}, {\it Ámbito Financiero}, {\it TN Noticias} and {\it El Destape Web}.

\par In Argentina, a pronounced polarization has been reported through the distinct ideological orientations of the country's primary media outlets \cite{digitalnewsreport2023,cicchini2022news}. For instance, Página 12 is recognized as a left-of-center broadsheet newspaper, while Clarín is considered a centrist tabloid and La Nación is characterized as a center-right newspaper \cite{bonner2018}. Between 2008 and 2014, a confrontation occurred between the government of Cristina Fernández de Kirchner (Center-Left) and major media corporations \cite{mitchlstein2017}. During this period, a conflict arose, leading to the establishment of a set of newspapers aligned with the policies of the Kirchner government (e.g., Página 12). Simultaneously, another cluster of newspapers emerged, known for their vehement editorial criticism of the government's actions during this era (e.g., Clarín and La Nación, among others) \cite{mitchlstein2017, Becerra2012, Yeager2014}.

\par While the examination of media outlet bias is increasing, particularly among English-based outlets, the scenario is different in countries like Argentina. Notably, only  three of the main outlets in Argentina have definitive bias classifications provided by Media Bias/Fact Check \cite{mediabiasfactcheck_clarin,mediabiasfactcheck_lanacion,mediabiasfactcheck_infobae}. In this context, our study not only classified Argentine news outlets but also introduced a versatile bias index for situations where specific classifications are lacking. 

\section{Material and Methods} \label{method}

\par
\begin{figure}
    \centering
    \includegraphics[width=0.8\textwidth]{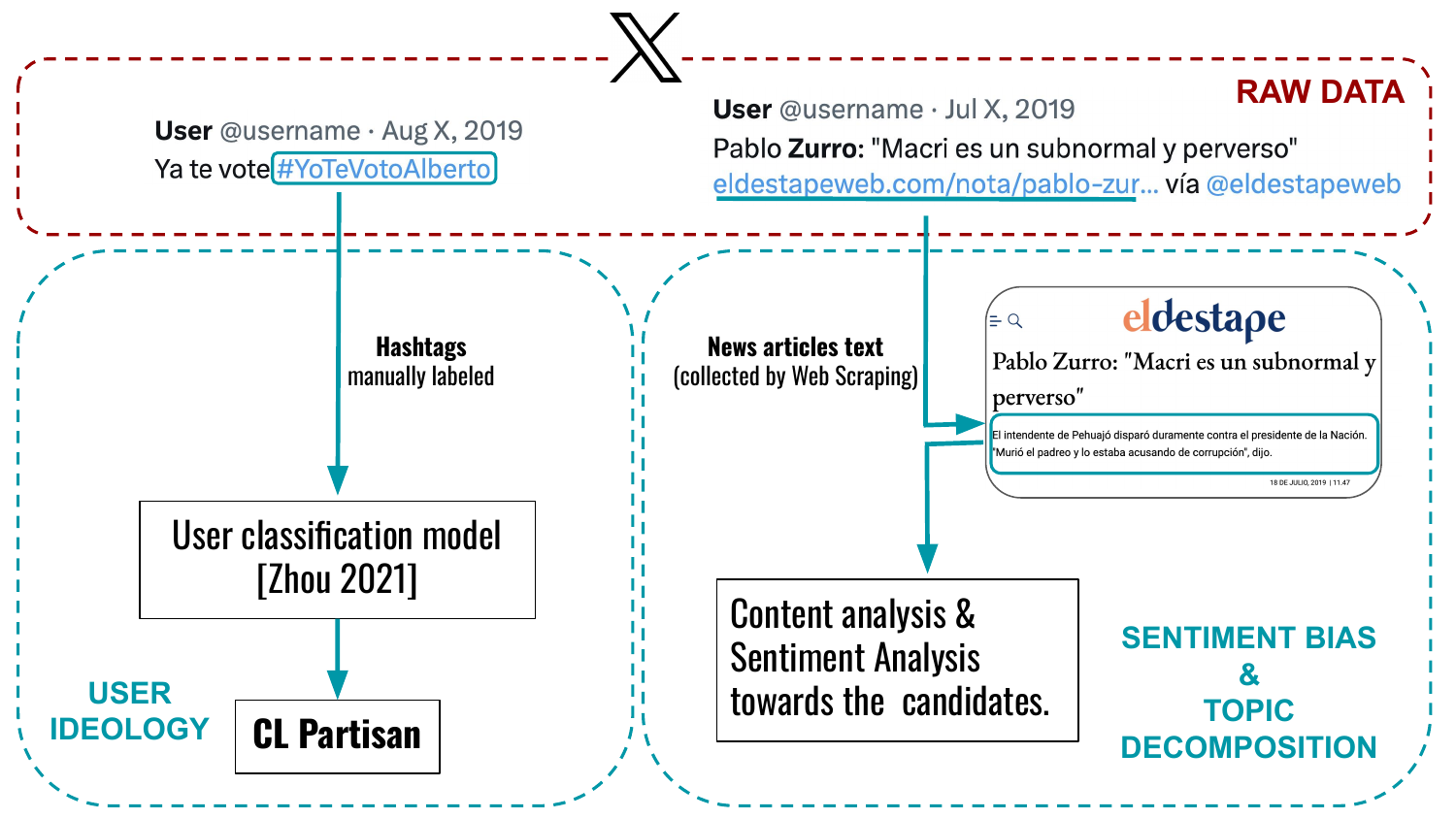}
    \caption{\textbf{Methodology pipeline.} \textbf{Top:} example of raw data of tweets from social media Twitter (now X). \textbf{In the left:} hashtags were utilized to train a logistic regression model for classifying tweets as supportive of one candidate or the other. Users are assigned to the candidate for whom they exhibit the highest number of supportive tweets (see more details in \cite{zhou2021}). \textbf{In the right:} the news URLs in the tweets are utilized to extract the text by web scrapping to execute then all the necessary steps leading to perform sentiment and topic analysis.}
    \label{fig:methodpipeline}
\end{figure}

 This section provides an overview of the data and methods utilized in this study. Figure \ref{fig:methodpipeline} illustrates the progression of our pipelines, commencing with the raw tweets (top panel of Fig. \ref{fig:methodpipeline}). Users are classified as supporters of a particular candidate based on the content of their tweets \cite{zhou2021}. Additionally, tweets containing URLs to external media outlets undergo scraping (right panel of Fig. \ref{fig:methodpipeline}), allowing for the analysis of news outlet bias, news, and topic bias based on the text of the news, rather than the text of the tweets themselves. Below is a detailed description of our methods.

\subsection{Users' classification}

A manually classified set of hashtags is used to construct a training set, which
was then employed to train a classifier (depicted in the left panel of Fig. \ref{fig:methodpipeline}).
This classifier categorizes tweets with a p-value ranging between 0 (pro-Fernandez, CL) and 1 (pro-Macri, CR). After classifying all the tweets from each user, users are ultimately assigned to the candidate
for whom they exhibit the highest number of supportive tweets. It's important to note that both
in the training process and in the classification process, the model employed in \cite{zhou2021}
only considers the text of the tweet; any external information contained in the tweet,
such as references to a news outlet, is not taken into account. Further details can be
found in \cite{zhou2021}.

\subsection{Data}

This study starts with an existing Twitter dataset \cite{zhou2021} containing
tweets collected between March 1, 2019, and August 27, 2019. The data was obtained
using keywords associated with candidates for the 2019 Argentina primary election,
including alferdez, CFK, CFKArgentina, Kirchner, mauriciomacri, Macri, and Pichetto. 

We refined the original dataset by a) including only tweets containing an external URL linking to an Argentinian news outlet and b) considering users involved in computing the final vote intention. This process yielded 65,971 tweets from 17,466 users intending to vote for the Center-Left (CL) coalition (Fernández-Fernández) and approximately 40,211 tweets from 15,425 users intending to vote for the Center-Right (CR) coalition (Macri-Pichetto). Intending CL coalition voters shared 19,395 news articles, while intending CR voters shared 10,219. The tweets considered in this work represent approximately $0.1\%$ of the raw data (see Supplementary Information of \cite{zhou2021}).

It's noteworthy that the while users' political orientation
was computed in \cite{zhou2021} by considering all the tweets of
a user (with and without a URL), and the model was trained using  a set of hashtags,
in this paper, we concentrate on a subset of those tweets (those containing a URL) and on the text of the news articles themselves, which was not utilized in \cite{zhou2021}.

\subsection{Data filtering} 

In order to acquire the primary dataset for our analysis, we implemented the following procedures:
\begin{enumerate}
    \item \textbf{Tweets with shared news selection:} We filter all tweets from the data collected by Zhou et al. (2021) \cite{zhou2021} that contained a URL in the \textit{url$\_$expanded} Twitter field. This included tweets, retweets, and quotes.
    \item \textbf{Urls expansion:} Requests python library \cite{requests} is used to expand the urls, applying multiprocessing.Pool.map() \cite{multipro} to parallelize the process.
    \item \textbf{Urls filter by media:} We retain only the URLs corresponding to news from Argentine media outlets based on
 ABYZ News Links Guide
  \cite{abyz}.
    \item \textbf{Scraping news articles:} For each media outlet, we develop a dedicated code to scrape the content from their respective web pages based on the python libraries Requests \cite{requests}, Selenium \cite{selenium} and Beautiful Soup \cite{bs4}. We acquire the texts of the news articles shared by users.
\end{enumerate}

\subsection{News Articles Sentiment Analysis}

\par After scraping, we perform sentiment analysis on the text of the shared news articles. We decompose each article into multiple sentences and apply Pysentimiento algorithm \cite{pysentimiento}  to each sentence within every article. This allows us to calculate positivity, neutrality, and negativity levels with regard to the two main election candidates. Sentiment is defined only for sentences that mention the candidates. If there is a single mention, it is counted as one. If there are multiple mentions, sentiment is calculated separately for each mention, categorizing them as neutral, positive, or negative. 

\subsubsection{Sentiment Bias}

\par We define the Sentiment Bias (\textit{SB}) \cite{albanese2020analyzing, cicchini2022news} of a news article as the balance between positive and negative mentions of the  candidates of the \textit{CL} coalition (Fernández-Fernández) versus the  candidates of the \textit{CR} coalition (Macri-Pichetto) using the following formula:
\begin{equation}
SB = \frac{(\#CR_{+}- \#CR_{-})-(\#CL_{+}- \#CL_{-})}{\#CR_{total} + \#CL_{total}}
    \label{sb_definition}
\end{equation}
where each mention is defined per sentence and the total number of mentions counts positive, negative and neutral ones.

For example, if an article has six sentences with mentions to candidates: one negative mention of CR candidates ($\#CR_{-}=1$), two positive mention of CL candidates ($\#CL_{+}=2$), and three neutral mention to CL candidates, then $\#CR_{+} = 0$, $\#CL_{-} = 0$, $\#CR_{total}=1$ and $\#CL_{total}=5$. The Sentiment Bias of the article is calculated as calculate $SB=\frac{(0 - 1)-(2 - 0)}{1 + 5} = \frac{-3}{6}=-0.5$. 

\subsubsection{Interpretation of the Sentiment Bias}\label{sec:interpretation}

\par Since Sentiment Bias ($SB$) is a fundamental metric in this study, this section delves into its analysis and provides a detailed interpretation. To conduct this analysis, we first manually classified a group of articles by selecting a random sample of 120 articles with well-defined SB, that is, articles in which candidates from either the Center-Left (CL) or Center-Right (CR) coalitions are mentioned. We then applied the majority rule to this manual classifications to obtain a unique label for each coalition. For instance, if an article received classifications of two positive, two negative, and two neutral with respect to a given coalition, we labeled the article as neutral for that coalition. Finally, we determined the overall connotation of the article based on the following criteria:

\begin{center}
\renewcommand{\arraystretch}{1.2} 
\setlength{\tabcolsep}{10pt} 

\begin{tabular}{|c | c | c|} 
 \hline
 Connotation over CL & Connotation over CR & Overall connotation \\ 
 \hline
 -1 & -1 & Neutral (0) \\ 
 \hline
 -1 & 0 & Favorable CR (1) \\
 \hline
 -1 & 1 & Favorable CR \\
 \hline
 0 & -1 & Favorable CL (-1) \\
 \hline
 0 & 0 & Neutral \\
 \hline
 0 & 1 & Favorable CR \\
 \hline
 1 & -1 & Favorable CL \\
 \hline
 1 & 0 & Favorable CL \\
 \hline
 1 & 1 & Neutral \\ 
 \hline
\end{tabular}
\end{center}

\par Given the overall connotation of each article, we applied logistic regression to correlate the SB value assigned to an article with its label. Specifically, we propose:

\begin{equation*}
P(l = i | SB) = \frac{e^{a_i SB}}{\sum_i e^{a_i SB}}
\end{equation*}

where $l$ is the connotation of the article, and $i = -1, 0, 1$ represents being favorable towards CL, neutral, and favorable towards CR, respectively. The coefficients $a_i$ are inferred by fitting the model to the labeled data. In order to keep the model as simple as possible, we chose not to include intercepts $b_i$ in the exponent of the exponential functions (i.e., $a_i SB + b_i$), after finding them to be insignificantly different from zero. The estimated coefficients are as follows: $a_{-1} = -0.89$ $[-1.44, -0.44]$, $a_{0} = -0.37$ $[-0.82, 0.07]$, and $a_{1} = 1.26$ $[0.82, 1.94]$. The numbers in brackets denote the 90\% confidence intervals, which were calculated using bootstrapping.

\par
\begin{figure}
    \centering    
    \includegraphics[width = \columnwidth]{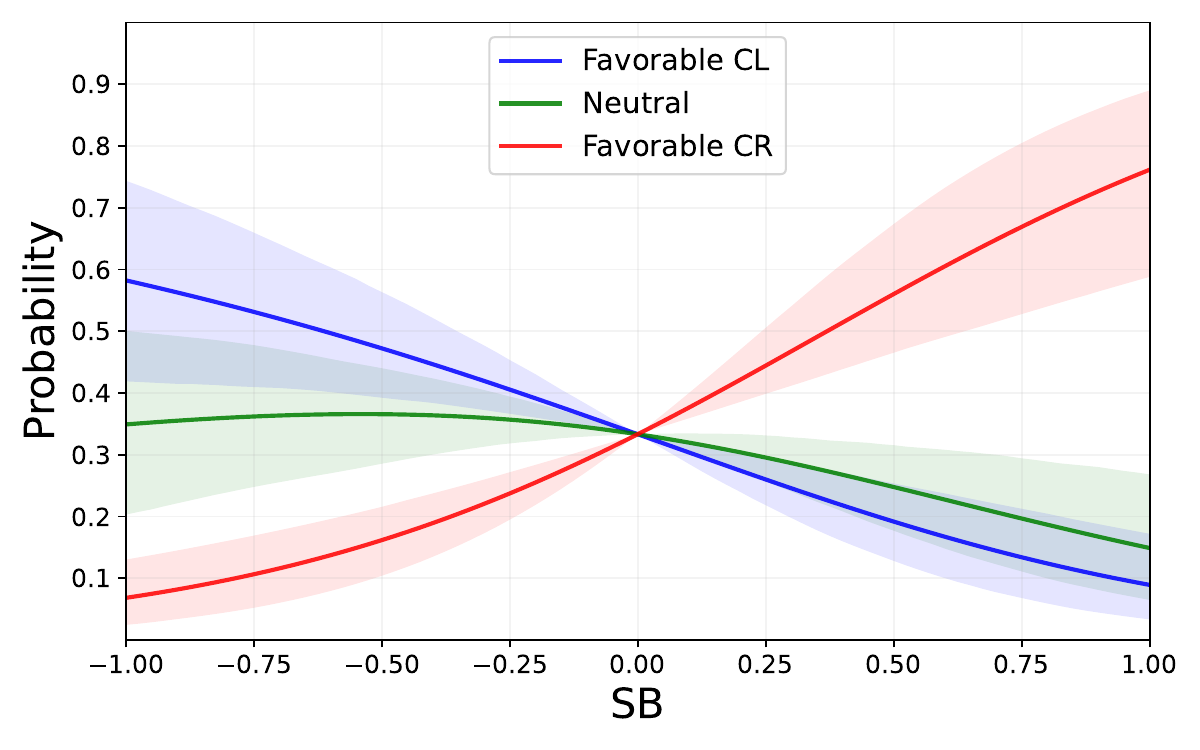}
  \caption{\textbf{Interpretation of the SB.} This figure displays the probability of an article being favorable towards CL, CR, or neutral, given the value of SB measured over the article content. The shaded regions represent the 90\% confidence intervals calculated by bootstrapping.}
    \label{fig:sb_validation}
\end{figure}

 In Figure \ref{fig:sb_validation}, we present the inferred probability of an article's connotation based on the measured value of $SB$. This figure facilitates the interpretation of the $SB$ value. For instance, a $SB = 0$ indicates an equal probability for an article to be either neutral or positive towards a given coalition. An article with a $SB$ slightly deviating from zero already indicates a clearly favorable trend towards a specific coalition. On the other hand, extreme values ($SB = -1$ or $SB = 1$) do not necessarily represent a probability equal to 1 of being favorable to a certain coalition. Instead, there is a significant fraction of neutral articles with these $SB$ values, and a small fraction of articles that express the opposite opinion, likely due to misclassifications by the sentiment detection algorithm \cite{pysentimiento}. Additionally, we observed a slight asymmetry for extreme $SB$ values, with a higher probability of an article being neutral when $SB = -1$ compared to when $SB = 1$.

\subsection{Topic decomposition}

\par We process the content of the articles by describing the texts within the bag-of-words framework. Specifically, we represent the corpus as a matrix of documents and terms, allowing subsequent topic description. To do this, we proceed with the following steps: 
    
    \begin{itemize}
        \item \textbf{Pre-Processing Text.}

    Given that we use a text representation based on word frequency, it is important to delete, on one hand, those words that are redundant and, on the other hand, those words that are non-informative, such prepositions and articles, in order to represent texts on a reduced set of meaningful words. This set of words will constitute our "vocabulary".
    
    \par With this in mind, we perform two things: First, we apply lemmatization on the texts using the python library Spacy \cite{spacy}, specifically we use \textit{es$\_$core$\_$news$\_$md} model \cite{lemma_model}. 
    Lemmatization transforms all the words to their roots, for instance, all verbs are transformed to their infinitive form and all substantives are transformed to their singular form.
    Then, we remove stopwords defined in NLTK python library \cite{nltk} (which, for instance, includes articles and prepositions), as well as rare words (that we defined as those that appears  in only one text of the corpus) and very frequent words that were not included in the stopword list but were present in more than 70$\%$ of the news articles.
    \end{itemize}
    \begin{itemize}
    
        \item   \textbf{TF-IDF}
    
    After defining the vocabulary, we proceed to describe texts in the bag-of-words framework.

    We start by describing each article by a term-frequency (TF) vector. This description transforms a given text to a vector where each component points out the number of times a given word of the vocabulary appears in the text. 
    We construct this representation through the object \textit{CountVectorizer} from the python scikit-learn library \cite{sklearn}.
    
    Moreover, to reduce word frequency bias and boost the impact of meaningful words, we compute for each word the Inverse Document Frequency (IDF) coeficient, defined as $\text{idf}_j = log(\frac{N}{N_j})$, where $N$ represents the total number of articles within the corpus, while $N_j$ denotes the count of articles containing the $j$-th term. To do this calculation, we apply the object \textit{TfidfTransformer} from \cite{sklearn}.

    With these ingredients, each text is finally described with the Term Frequency - Inverse Document Frequency (TF-IDF) coeficients \cite{NGUYEN201495}, where the $j$-th component of the "article vector" $i$ is given by:
\begin{equation*}
        v_{ij} = f_{ij} \cdot \log \left(\frac{N}{N_{j}}\right)
\end{equation*}
    where $f_{ij}$ is the frequency of term $j$ in article $i$ and $N_j$ is the number of documents where the term $j$ appears, as it was stated before.
    
    Then the articles corpus is described as a matrix $M \in \mathrm{R}^{n \times m}$, with $n$ the number of articles in the corpus and $m$ the number of terms included in the vocabulary. This matrix is a concise representation of the corpus where the meaningful words (both frequent and specific words) are enhanced for each text.

      \item   \textbf{Topic Decomposition.}
 
In this step, in order to identify the main topics of the corpus of news articles, we apply the unsupervised topic detection algorithm Non-negative Matrix Factorization (NMF) model from scikit-learn python library \cite{sklearn} on the news-term matrix $M$ constructed in the previous step. NMF decomposes matrix M into the product of two matrices, ensuring that all elements are non-negative: 

\begin{equation*}
M \approx H \cdot W \textit{ , where } H \in \mathrm{R}^{n \times t}\textit{ and }W \in \mathrm{R}^{t \times m}
\end{equation*}

Here, $t$ represents the selected number of topics, and $H$ and $W$ denote the resulting matrices of the decomposition. In particular, $H$ defines how each article is described in terms of topics. The element $h_{ij}$ points out the weight of topic $j$ on article $i$. In other words, it quantifies how much article $i$ belongs to topic $j$. In order to interpret these weights in terms of probabilities, each row is normalized such as $\sum_j^t h_{ij} = 1$.

On the other hand, rows of matrix $W$ specify the description of each topic in terms of the vocabulary built above. In this case, the element $w_{ij}$ denotes the weight of term $j$ in topic $i$, i.e, how well term $j$ describes topic $i$. In this case, by only identifying the weightiest terms allows to interpret what the topic talks about.

\end{itemize}

\subsubsection{Media agenda}
 
 Following the procedure outlined in \cite{PaperSeba}, we define the media agenda as the proportion of articles associated with each topic.
 Specifically, we define the weight of topic $j$, $T^j$, as:
 \begin{equation}
     T^j = \frac{1}{n} \sum_i^n h_{ij}
     \label{eq:topic_definition}
 \end{equation}
with $h^{ij}$ being the weight of topic $j$ on article $i$ (as defined earlier) and $n$ the number of unique articles shared by the media outlets. We interpret  $T^j$ as the collective interest of media outlets in topic $j$. This measure indicates the likelihood of finding an article associated with topic $j$ in our dataset. (in this case, we are not considering the number of times each article was sharing in social media. Therefore, this definition holds for unique articles).

 \subsubsection{Partisans agenda}
 
In order to distinguish the interest of partisans groups over the topics found above, we define the interest of partisan group $p$ over topic $j$ as the average of elements $h_{ij}$ (weight of topic $j$ on article $i$) weighted by the number of times group $p$ shares article $i$ ($s_{pi}$):
  \begin{equation}
     T^j_p = \frac{\sum_i^n s_{pi} h_{ij}}{\sum_i^n s_{pi}}
     \label{eq:topic_definition_partisans}
 \end{equation}
 where $\sum_i^n s_{pi}$ is equal to the total number of times users from group $p$ shared an article $i$ and $n$ being the total number of articles.
 $T^j_p$ tells us the probability that an article associated with topic $j$ is shared by an user identified with group $p$.

\section{Results}

\par As outlined in the Introduction, the aim of this study is to investigate how and which characteristics of shared news, and to what extent, correlate with the political ideologies of the users sharing them. To achieve this, we analyze various characteristics of news articles shared by users with identified political leaning, encompassing their sources, distribution of topics, and the political biases that manifest at several levels.

\subsection{Data description}

\par
\begin{figure}[htbp]
    \centering
    \includegraphics[width=0.75\textwidth]{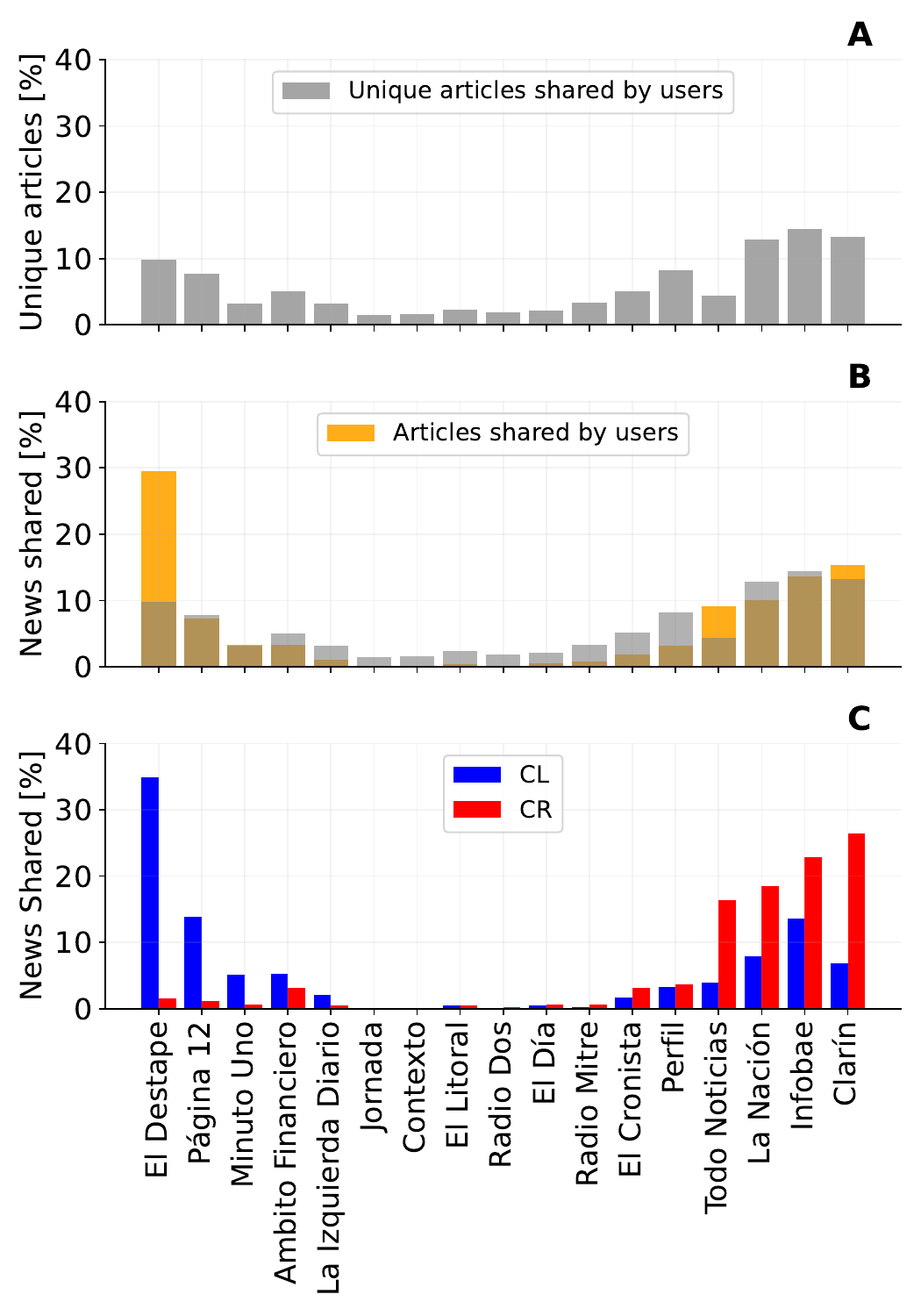}
  \caption{\textbf{Distribution of News Articles by Media Outlet.} \textbf{(A)} Shown in gray, the distribution of unique news articles circulating on Twitter, categorized by their source media outlet (\textit{unique} indicates that the frequency of news sharing is not considered). \textbf{(B)} Illustrated in yellow, the distribution of news articles shared by users, with each instance of news sharing counted independently, regardless of previous shares by others (in gray, the distribution from panel A for comparison). \textbf{(C)} The percentage distribution of news articles shared by media outlets among CL (in blue) and CR (in red) supporters, based on their sharing patterns.
    \label{news_by_media}}
\end{figure}

 Figure \ref{news_by_media} A shows the distribution of news articles on Twitter, categorized by their originating media outlets. By \textit{unique} we imply that each article is counted only once, no matter how many times was posted. This diagram illustrates that approximately 60\% of the articles originate from a set of specific outlets: Infobae, Clarín, La Nación, El Destape, Perfil, and Página 12, listed in descending order by count. This distribution mirrors the activity level of these outlets, with Infobae being the most active in terms of articles published.

\par Figure \ref{news_by_media} B depicts the distribution of news articles shared by Twitter users, taking into account the frequency of each article's sharing. This highlights the impact of user preferences on the distribution observed in panel A. For instance, articles from El Destape constitute about 30\% of the shared content, underscoring its significance despite not being the highest in publication volume. The same six outlets (with Perfil replaced by Todo Noticias) account for approximately 80\% of the shared articles. We juxtapose this data with the information from panel A (represented by gray bars) for a comparative analysis.

\par An interesting pattern emerges upon examining users' ideological orientations. Referencing the work by Zhou et al. \cite{zhou2021}, we categorized users' political biases towards either the Center-Left (CL) or Center-Right (CR) factions during the 2019 Argentine presidential elections. Figure \ref{news_by_media} C shows the distribution of articles shared by each group, analyzed separately. CR supporters predominantly share content from Clarín, La Nación, Todo Noticias, and Infobae, whereas CL supporters mainly distribute articles from El Destape and Página 12, which are largely overlooked by CR adherents.

\par While associating specific media outlets with particular political biases might seem straightforward, our analysis reveals that CL supporters also engage significantly with content from outlets viewed as CR-aligned in panel C of Fig. \ref{news_by_media}, such as Clarín, La Nación, Todo Noticias, and notably, Infobae. Conversely, external evaluations classify Clarín and La Nación as Center-Right biased, with Infobae showing a Center-Left inclination \cite{mediabiasfactcheck}. Hence, the correlation between the political leanings of social media users and their shared media sources is more nuanced. The following sections will explore this relationship with more detail.

\subsection{Sentiment Bias}

\par To better understand how users' political ideology relates to the news they share on social media, we'll start by examining the bias of the news articles. Bias refers to the tendency of an article to lean positively or negatively towards one of two political coalitions, CL and CR. To analyze bias, we use the Sentiment Bias ($SB$) metric introduced in section \ref{method}. Essentially, $SB$ gives us a score between $-1$ and $1$ for each article mentioning a candidate from either coalition. A score closer to $-1$ indicates a favorable stance towards CL, while a score closer to $1$ indicates a favorable stance towards CR. This metric helps us define the bias of each article and, consequently, of each media outlet.

\par
\begin{figure}[htbp]   
    \centering
    \includegraphics[width=0.95\textwidth]{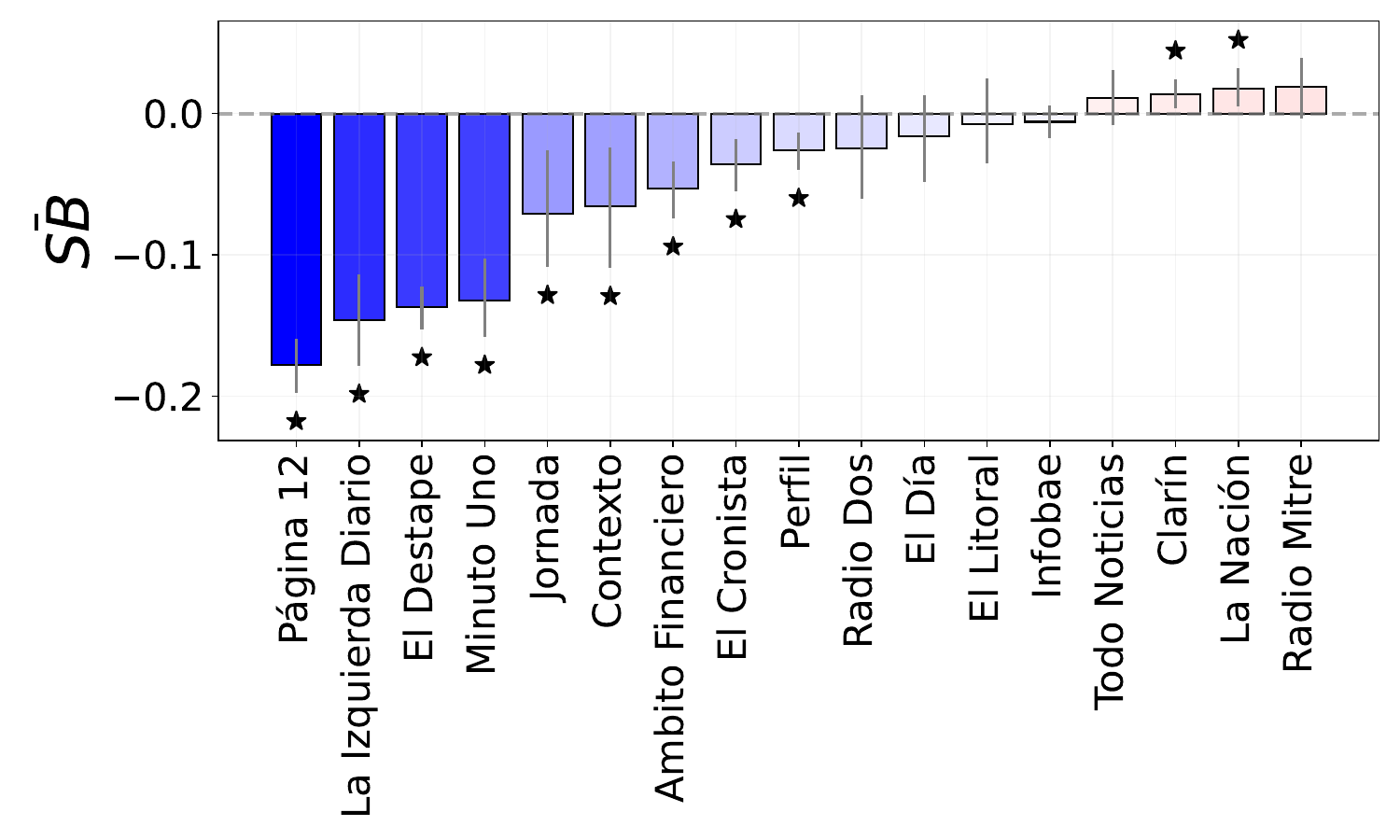}
    \caption{
    \textbf{Mean Sentiment Bias ($\Bar{SB}$) by media outlet.} $\Bar{SB}$ represents the average SB across all articles from a specific media outlet. Media outlets positioned on the left side are interpreted as having a bias towards the Center-Left (CL), while those on the right side are considered to have a bias towards the Center-Right (CR).
    Gray bars indicate the centered $99\%$ quantile of the estimator determined through bootstrapping, and stars denote those estimates significantly different from zero.}
    \label{sb_by_media}
\end{figure}

 Figure \ref{sb_by_media} shows the average sentiment bias ($\bar{SB}$) for each media outlet, which is the average of the $SB$ scores of all articles from that outlet. For instance, Página 12 and El Destape exhibit $\bar{SB}$ values favoring CL, whereas La Nación and Clarín show $\bar{SB}$ values favoring CR. Notably, Infobae, shared by both supporter groups, falls between these two groups of outlets.
Regarding the absolute value of $\bar{SB}$, we interpret $\bar{SB} = 0$ as a neutral position (see section \ref{sec:interpretation}), meaning most outlets slightly favor CL during the analyzed period. 
El Destape and Página 12 are more extreme in their positions and can be certainly considered as Center-Left outlets, while Clarín, and La Nación, closer to the center, can be also considered centrist media but slightly lean towards the Center-Right position.
For all mentioned media outlets, $\bar{SB}$ significantly deviates from zero, unlike the case for Infobae, which underscores its apparent centrist position.

\subsubsection{Selective sharing}

\par
\begin{figure}[htbp]   
    \centering
    \includegraphics[width=\textwidth]{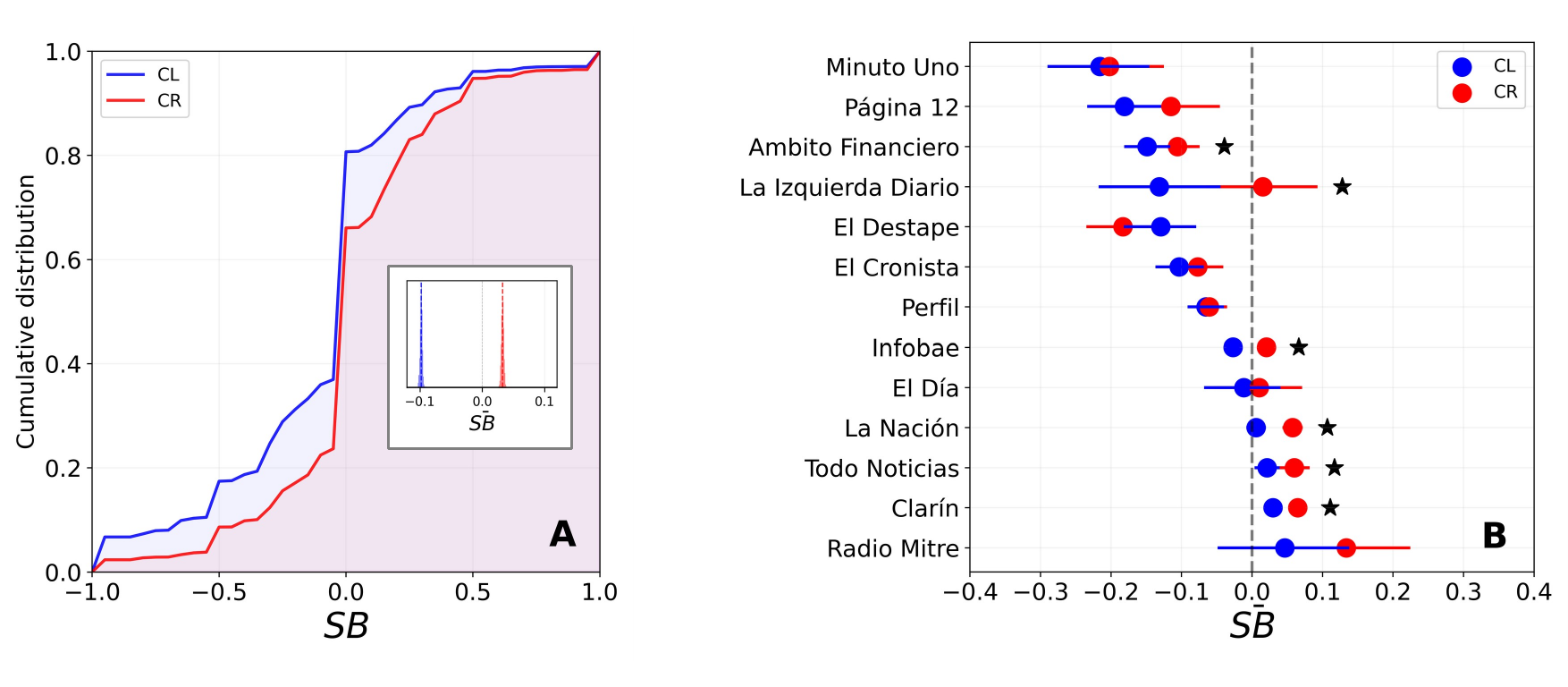}
    \caption{
\textbf{(A) Cumulative Distribution of Sentiment Bias (SB) for shared news articles.}  
Inset: average Sentiment Bias for news articles shared by CL and CR Partisans.
The distributions illustrate the variability in the estimates calculated through bootstrapping.
\textbf{(B) Average Sentiment Bias for news articles shared by CL and CR partisans across media outlets.} Each point represents the average sentiment bias across all articles from a given media outlet shared by each user group, with CL in blue and CR in red. Only media outlets with at least 100 articles shared by each group are included (see Supplementary Information).
The horizontal bars indicate the centered $99\%$ quantile of the estimate obtained via bootstrapping. Black stars highlight instances where the difference in $\bar{SB}$ between CL and CR supporters is statistically significant, with $p < 0.01$.}
    \label{sb_partisans_media}
\end{figure}

 Having determined the bias of the articles, we analyze which ones were shared on social media by users with clearly identified political positions.
Panel A of Fig. \ref{sb_partisans_media} presents the cumulative distribution of the Sentiment Bias ($SB$) of news articles shared by users identified with one of the two political coalitions. This panel illustrates a clear tendency: the $SB$ of articles shared by Center-Left (CL) supporters shifts towards negative values, indicating a preference for sharing articles that favor their coalition. Similarly, for Center-Right (CR) supporters, the distribution of $SB$ skews positively (we reject the hypothesis that the $SB$ distribution for CR supporters exceeds that for CL supporters with $p < 10^{-3}$, using a Kolmogorov-Smirnov test). The inset of this panel depicts the estimated mean value $\bar{SB}$
of these distributions.
CL voters generally share news articles with an average $\bar{SB}$ of approximately $-0.10$, whereas CR voters share articles with an average $\bar{SB}$ of approximately $0.03$, with both averages being significantly different from zero.

\par What is most interesting is when we break down the previous analysis by media outlet. As discussed in panel C of Fig. \ref{news_by_media}, although certain groups of media outlets tend to be more shared by each political coalition, there is a subset (such as Clarín, La Nación, and Infobae) which is significantly shared by both coalitions. However, the content extracted by each coalition from these outlets differs.
Panel B of Fig. \ref{sb_partisans_media} displays the average $SB$ of articles shared by each group, now categorized by outlet. This illustrates the phenomenon known as ``cherry picking", where users share news that align with their political beliefs, even from opposing outlets. For example, left-leaning supporters share news from La Nación (identified as a right-biased news outlet in Fig. \ref{sb_by_media}) with an average $SB$ close to zero, while right-leaning supporters share news from the same outlet with a higher average $SB$. Similar trends are observed in Clarín (Center-Right), Infobae (Centrist), and La Izquierda Diario and Ámbito Financiero (Center-Left). We have statistically validated significant differences between the groups for each outlet, with a p-value below 0.01. Statistically significant differences in news sharing biases are marked with a star in Fig. \ref{sb_partisans_media}. 

\subsection{Topics interest}

\par This section delves into whether users' political inclinations also affect the topics of the news they share. The findings in previous section establish a link between the users' ideologies and the political bias in the news they share. Here, we aim to determine whether specific themes are more supportive of particular candidates and if supporters of each coalition show a preference for these topics.

\par We initially conduct a topic decomposition of the news articles to identify the principal themes within the dataset, as detailed in section \ref{method}. We identified two main families of topics: the first related to economic issues, such as \textit{Wage/Inflation} and \textit{Economy/Dollar}; the second pertains to topics associated with the presidential elections occurring during the analyzed period, including \textit{Politics CR}, \textit{Politics BA Province}, \textit{3rd Party}, \textit{Elections}, \textit{Politics CL}, and \textit{Justice}.
Descriptions of these topics, including word clouds and examples of related news articles, are available in the Supplementary Information.

\par
\begin{figure}[htbp]   
    \centering
    \includegraphics[width=\textwidth]{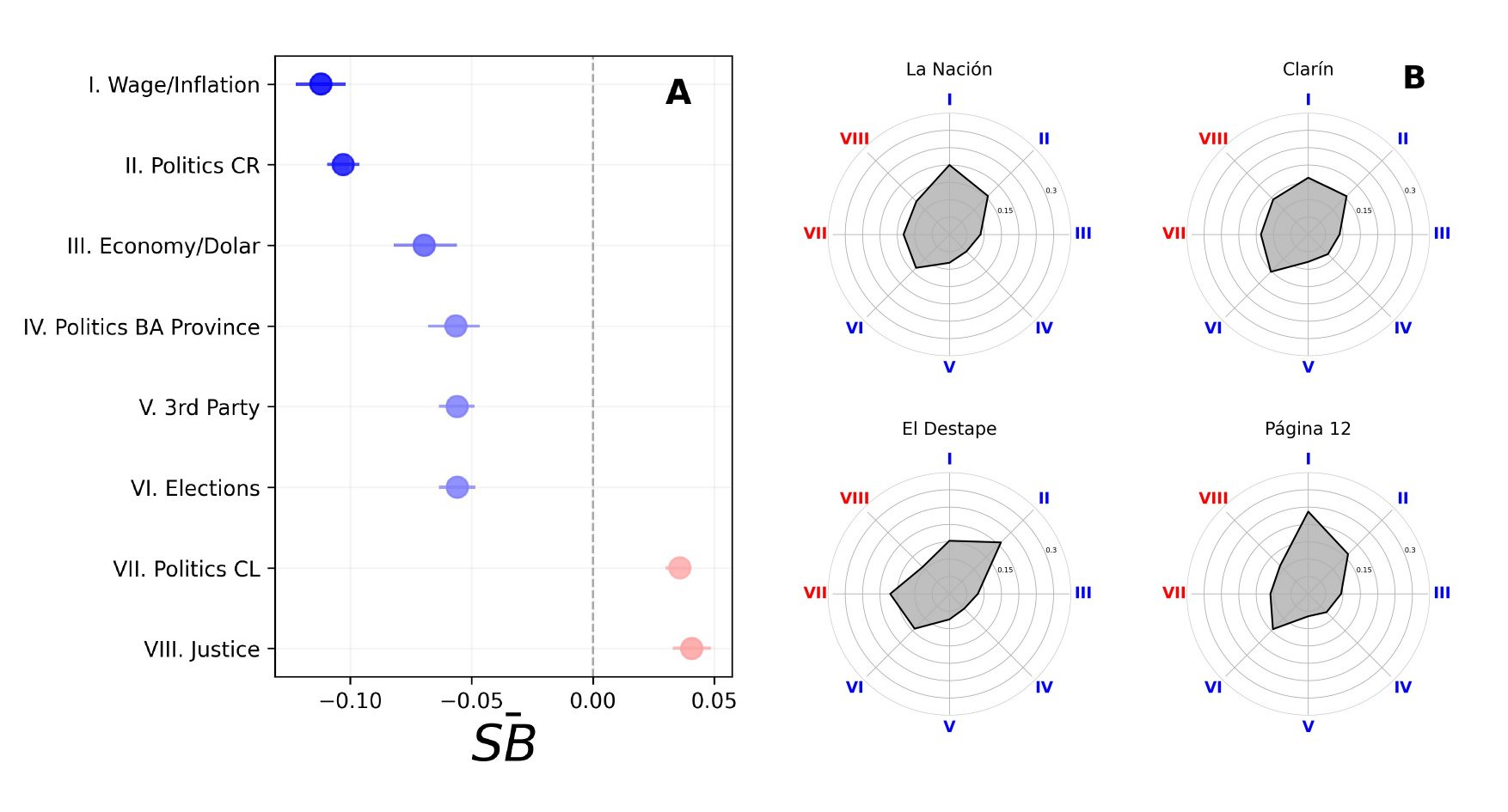}
    \caption{\textbf{(A) Average Sentiment Bias of emergent topics.} Topics are enumerated for subsequent reference. Colors indicate the sign of the $SB$ to highlight its orientation: blue signifies a positive bias towards CL, and red indicates a positive bias towards CR. \textbf{(B) Media Outlet Agendas.} Numbers correspond to the topics identified in panel A, with colors reflecting their respective biases. The outlets displayed are those with a clear leaning towards CR (La Nación and Clarín) and CL (El Destape and Página 12), as demonstrated in Fig. \ref{sb_by_media}.}
    \label{fig:media_agenda}
\end{figure}

 Regardless of the interpretation of these topics, which depends heavily on context, panel A of Fig. \ref{fig:media_agenda} provides insight into which topics are supportive or against each coalition by displaying the estimated $\bar{SB}$ for each topic. 
Given that each article is associated with each topic to a varying degree (refer to section \ref{method}), $\bar{SB}$ reflects the weighted average $SB$ of each article according to this association.
For example, this panel indicates that the topic \textit{Wage/Inflation} supports the CL stance, while \textit{Justice} leans towards CR. Notably, topics labeled as \textit{Politics CR} and \textit{Politics CL} appear to favor the coalition contrary to what their labels suggest, likely because they group articles critical of those coalitions.
The remaining topics exhibit a slight preference towards CL, aligning with the overall tendency observed during the analyzed period (refer to, for instance, Fig. \ref{sb_by_media}).

\par Furthermore, we explore the topics covered in news articles  to discern the ``agenda" of each media outlet (referred to as the ``media agenda" in section \ref{method}). This agenda essentially represents how each outlet distributes its coverage across the detected topics. Panel B of Fig. \ref{fig:media_agenda} showcases the agendas of four media outlets, 
two with a right-leaning bias (Clarín and La Nación) and two with a left-leaning bias (El Destape and Página 12). 
Upon inspecting this panel, clear similarities and differences emerge. 
Much of this coverage behavior can be understood by considering the overall bias of each topic as shown in panel A of Fig. \ref{fig:media_agenda} and the bias of each media as depicted in Fig. \ref{sb_by_media}.
For instance, Clarín and La Nación show a priority for covering \textit{Justice} compared to the other two outlets, whereas Página 12 exhibits a stronger focus on \textit{Wage/Inflation}, and El Destape on \textit{Politics CR}, relative to other topics.

\subsubsection{Partisans agenda}

\par
\begin{figure}[htbp]   
\includegraphics[width = \textwidth]{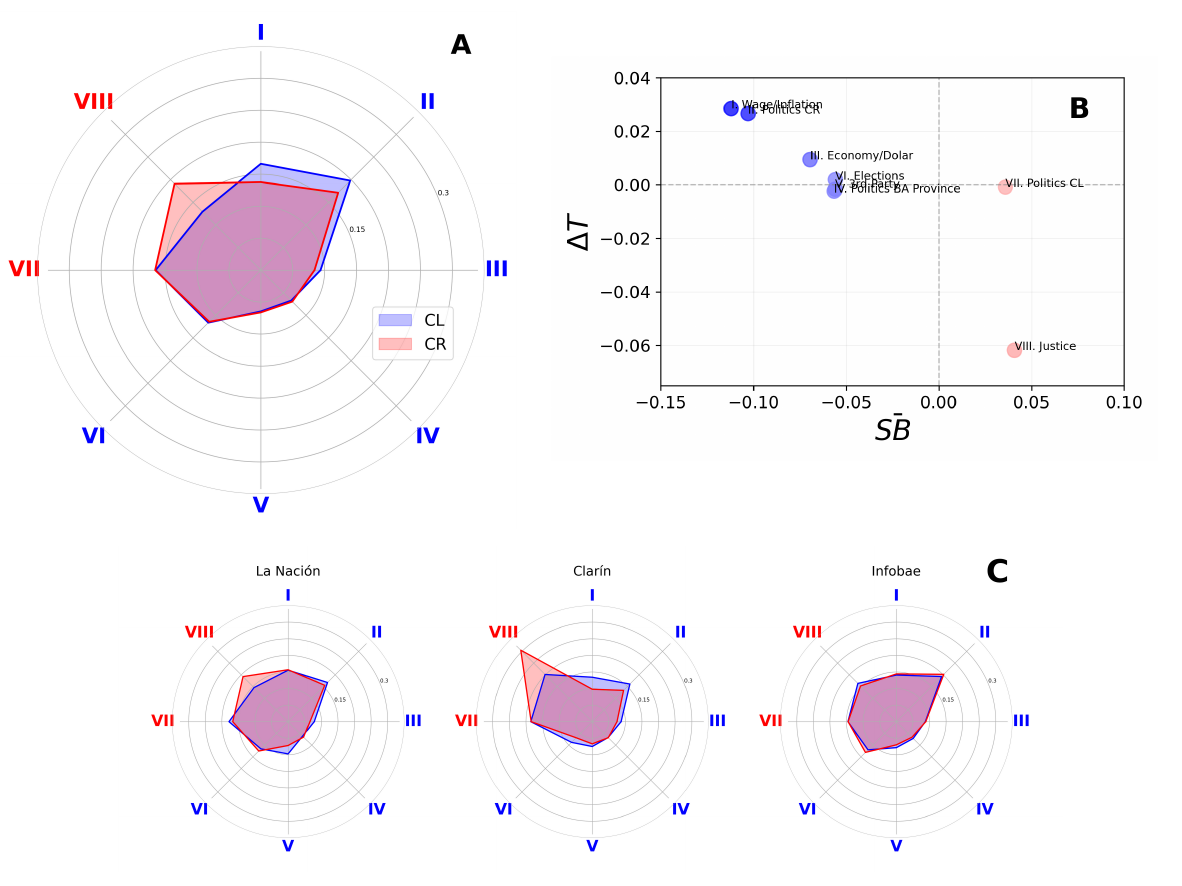}
    \caption{\textbf{(A) Partisan agendas.} Distribution of topic interests for each partisan group. Roman numerals refer to topics outlined in panel A of Fig. \ref{fig:media_agenda} and also in panel B of this figure.
    \textbf{(B) Difference in topic interests.} This difference is calculated using equation \ref{eq:topic_difference}.
    \textbf{(C) Partisan agendas by media outlet.} The outlets depicted are those significantly shared by both coalition groups (see Fig. \ref{news_by_media} C).}
    \label{fig:agenda_partisans}
\end{figure}

 The topics described above influence social media users according to their political leanings. These leanings may constrain users to prefer sharing certain topics over others. 
Panel A of Fig. \ref{fig:agenda_partisans} provides insights into the preferred topics for each partisan group, delineating what we term the ``partisan agendas". 
In this figure, it is evident that CL (Center-Left) users demonstrate a greater interest in topics like \textit{Politics CR} and \textit{Wage/Inflation}, which exhibits a positive inclination towards the CL coalition, whereas CR (Center-Right) users show a preference for the topic \textit{Justice}, with a positive bias towards the CR coalition. 
Panel B further clarifies the disparity in these interests.
We define this difference as
\begin{equation}
    \Delta {T}^j = T^j_{CL}-T^j_{CR}
    \label{eq:topic_difference}
\end{equation}
where $T^j_{CL}$ denotes the interest of CL partisans in topic $j$. 
As illustrated in panel B of Fig. \ref{fig:media_agenda}, the specified topics significantly align with the coalition of users who share them, indicated by a Spearman correlation coefficient of $-0.78$ (with a 90\% confidence interval of [-1, -0.24]). The sole noticeable exception is the topic \textit{Politics CL}, in which both coalitions appear to have an equal interest, yet it demonstrates an overall inclination towards CR.

\par Finally, panel C showcases the distribution of news shared by each partisan group, this time segmented by media outlet. 
This panel unveils another dimension of the cherry-picking behavior outlined in panel B of Fig. \ref{sb_partisans_media}.
For example, while users from both coalitions distribute news from Clarín and La Nación, CR users predominantly share content related to the topic \textit{Justice}, which exhibits a CR-favorable bias, whereas CL users are more inclined to share information on \textit{Wage/Inflation}, which aligns with a CL-favorable bias as indicated in panel A of Fig. \ref{fig:media_agenda}. Nevertheless, this cherry-picking behavior seems to be absent in the topic dissemination from the centrist outlet Infobae, in contrast to the observations made in panel B of Fig. \ref{sb_partisans_media}, where each group distinctly shared articles from this media outlet that were biased towards their respective preferences.

\section{Discussion and Conclusions}
In this work, we investigated the relation between shared news on social media and the ideologies of the users who share them. We analyzed both the source of the news articles, their intrinsic bias, and the topics covered, and we related each of these characteristics to the users political ideologies from \cite{zhou2021}. To accomplish this, we analyzed the content of news articles shared by politically aligned users on X (ex-Twitter), scraping their content and quantifying both the bias and the topics covered.

Our initial analysis focused on sources (i.e. media outlets). This analysis revealed that the sharing behavior of news by users did not exhibit a distinctly polarized distribution. While certain media outlets may be associated with particular political ideologies (CL and CR), we observed a significant percentage of news from Center-Right (CR) outlets being shared by users identified as Center-Left (CL). See Fig. \ref{news_by_media} C.  This suggests that the sources of news shared by users on social media may not necessarily indicate their ideology. Our data indicates that Center-Right (CR) media outlets are the most widely consumed in the country, aligning with findings from \cite{comscore}. Additionally, our results highlight that Center-Right (CR) media outlets reach a more diverse audience in terms of ideological spectra.

We delve deeper into the analysis of the relationship between users' ideologies and the news they share, by examining the bias of news content using the previously introduced Sentiment Bias index \cite{albanese2020analyzing, cicchini2022news}. This index effectively categorizes biases of news outlets without making any assumptions, as depicted in Fig. \ref{sb_by_media}. Our findings are consistent with external classifications, where available, validating the accuracy of our approach \cite{mediabiasfactcheck}. 

When analyzing the average Sentiment Bias ($\bar{SB}$) alongside social media data, a significant trend emerges: users on social platforms tend to share news that aligns with their political beliefs (Fig. \ref{sb_partisans_media} A). This tendency can be interpreted as indicative of the selective exposure theory \cite{stroud2010polarization}. The findings are supported by Fig. \ref{sb_partisans_media} B, confirming a ``cherry-picking" trend: users engage with various journals regardless of their political alignment, yet selectively choose news that resonates with their ideologies. This underscores their preference for content reinforcing their existing beliefs. Furthermore, our analysis extends these patterns to specific topics, as demonstrated in Fig. \ref{fig:agenda_partisans}. Users distinctly favor sharing articles related to subjects aligning with their preferred candidates.

While it's expected for users with defined ideological leanings to share news that aligns with their biases, the analysis presented here highlights that this tendency is only apparent when assessing the bias of the content itself, rather than solely relying on media bias. While this phenomenon is predictable, the aim of this study is to introduce a method for quantifying such behavior.

Finally, we'd like to address some remarks and potential limitations of our study. The dataset, while four years old and specific to Argentina, provides unique insights into users' political leanings not found in other datasets. User classification was achieved through a machine learning model, enhancing the dataset's value and enabling us to explore how it correlates with the political bias of shared news content. We believe this analytical framework could be valuable in other countries, especially those with pronounced political polarization like Argentina, and could be adapted to multipolarized scenarios \cite{martin2023multipolar}.

\section*{Acknowledgements}

HAM and MS were supported by NSF Grant No. 2214217. PB, SMdP, SP and LG were supported by PICT-2020-SERIEA-00966. 

\section*{Author contributions statement}

SMdP, SP, LG, and MS were responsible for collecting the raw data. SMdP developed the computational code utilized consistently throughout the paper. SMdP and SP made contributions to the statistical analysis. PB and HAM conceptualized the research. All authors engaged in discussions about the results and collaborated on the development of the manuscript.

\section*{Data availability}
The datasets generated and analysed during the current study are available in the OSF repository in \url{https://osf.io/sxwmj/} 
and the corresponding codes are avaiable in \url{https://github.com/sofiadelpozo/SocialMediaBiasAndPolarization}

\printbibliography



\section*{Supplementary material}

# Analyzing User Ideologies and Shared News During the 2019 Argentinian Elections.
# Supplementary Information

Sofía M del Pozo[1,2], Sebastián Pinto[1,2], Matteo Serafino[3], Lucio Garcia [1,2], Hernán A Makse[3] and Pablo Balenzuela[1,2].

[1] Universidad de Buenos Aires, Facultad de Ciencias Exactas y Naturales, Departamento de Física. Buenos Aires, Argentina.

[2] CONICET - Universidad de Buenos Aires, Instituto de Física Interdisciplinaria y Aplicada (INFINA). Buenos Aires, Argentina.

[3] Levich Institute and Physics Department, City College of New York, 10031 New York, USA.

## 1   News articles by partisans

Figure 1 shows the amount of news articles shared by users with Center-Left and Center-Right political leanings for each media outlet. Media outlets highlighted in red were excluded from the $\tilde{SB}$ analysis in Fig. 5 of the main text.

## 2   Topics of news articles

Figure 2 shows the wordclouds of the topics cited in the main text. Table 1 shows examples of related articles for each topic.

### Competing Interests Statement

The authors declare that they have no known competing financial interests or personal relationships that could have appeared to influence the work reported in this paper.

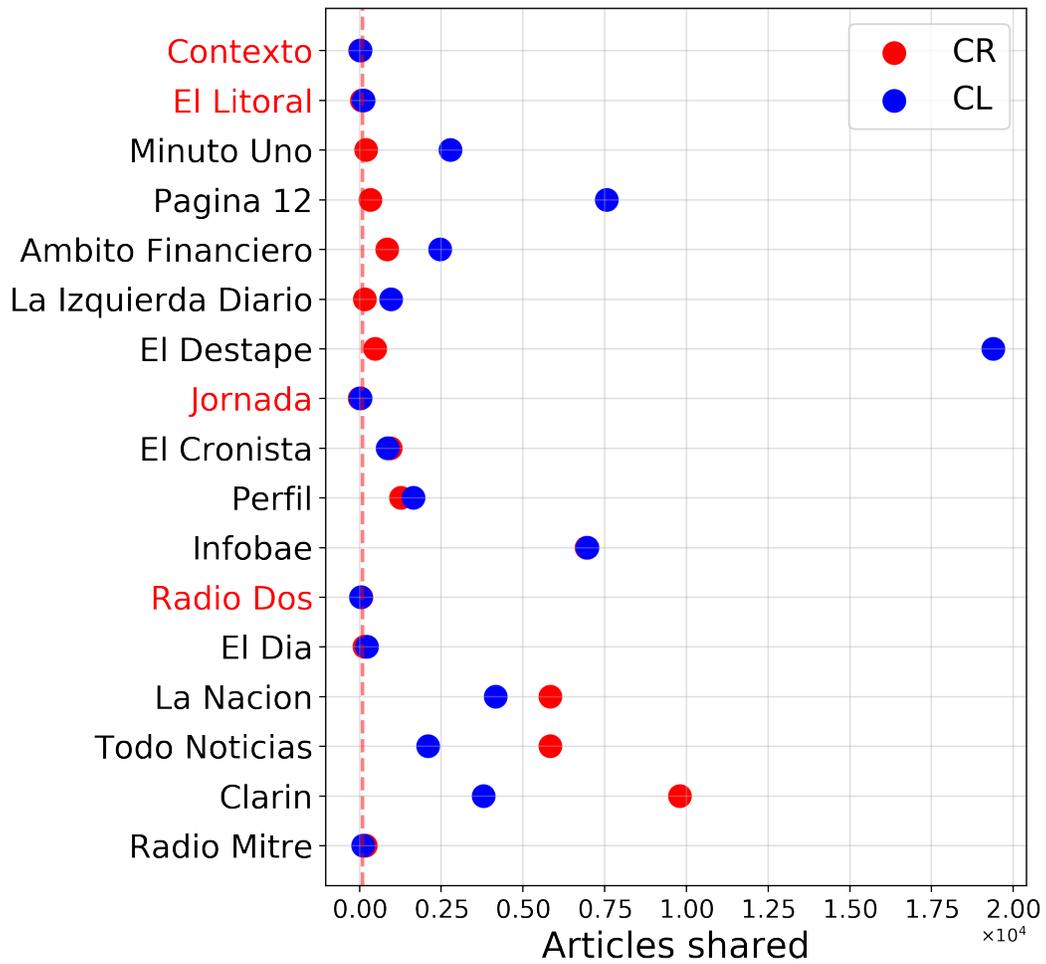

Figure 1: **Amount of news articles shared by users with Center-Left (blue) and Center-Right (red) leanings.** The red line represents the threshold of 100, which is when we began including media outlets for $\bar{SB}$ analysis in Fig. 5 in the main text. Media outlets in red have less than 100 news articles shared by at least one group of users.

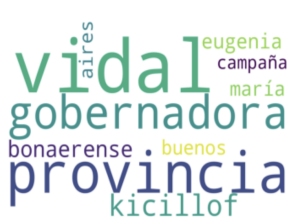

(a) Politics BA province.

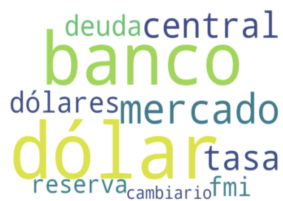

(b) Economy/Dolar.

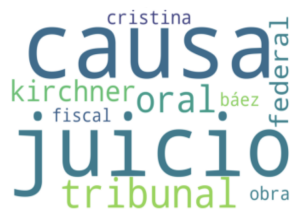

(c) Justice.

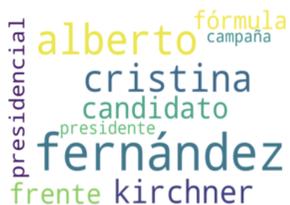

(d) Politics CL.

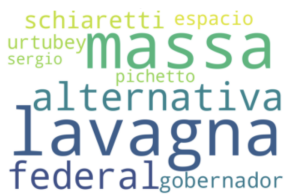

(e) 3rd Party.

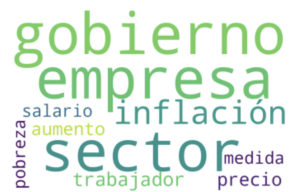

(f) Wage/Inflation.

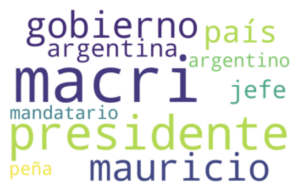

(g) Politics CR.

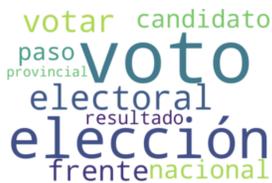

(h) Elections.

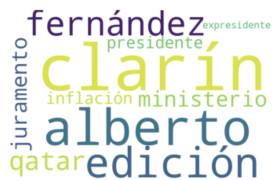

(i) Trash topic.

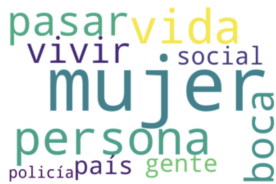

(j) Gossip & Sports.

Figure 2: Word clouds describing each topic.

3

| Topic | Title of news articles |
|---|---|
| Politics BA Province | El Destape: *Debate 2019: qué dijo María Eugenia Vidal tras el fallido de Macri en el que dio a Kicillof como gobernador* [1], <br> Página 12: *Un café amargo para María Eugenia Vidal* [2], <br> La Nación: *Así es la cumbia de Vidal que busca captar el voto bonaerense* [3] |
| Economy/Dolar | La Nación: *El dólar subió hoy 80 centavos por la pelea de Trump con China* [4], <br> El Destape: *Dólar hoy: subió más de 30 centavos y cerró a $ 40,13* [5], <br> Todo Noticias: *El Banco Central y las elecciones: dólar bajo control hasta las PASO y tasas altas por cinco meses* [6] |
| Justice | Infobae: *La Fiscalía solicitó unificar las causas "Hotesur" y "Los Sauces" y que Cristina Kirchner sea juzgada en un solo juicio oral* [7], <br> La Nación: *Cristina fue a Comodoro Py para notificarse de los nuevos procesamientos* [8], <br> Todo Noticias: *Lázaro Báez pidió suspender el juicio que tiene a Cristina como principal acusada* [9] |
| Politics CL | La Nación: *Los mejores memes por el anuncio de Cristina Kirchner como vicepresidenta de Alberto Fernández* [10], <br> El Destape: *Alberto Fernández le contestó a Cambiemos: "Cristina no es Perón y yo no soy Cámpora"* [11], <br> Página 12: *Romano: "La figura de Alberto Fernández es inspiradora"* [12] |
| 3rd Party | La Nación: *Lifschitz y Stolbizer evitaron una ruptura definitiva de Lavagna con Alternativa Federal* [13], <br> Todo Noticias: *Roberto Lavagna: "No hubo ninguna ruptura con Alternativa Federal, todos los diálogos están abiertos"* [14], <br> Página 12: *Tensión en el peronismo disidente por la estrategia electoral* [15] |
| Wage/Inflation | Clarín: *A marzo, los sueldos perdieron 10,5% de su poder adquisitivo* [16], <br> Página 12: *Fuerte retroceso de la industria* [17], <br> El Destape: *La eliminación del IVA beneficiará más a sectores de mayor ingreso que a los más pobres* [18] |
| Politics CR | Página 12: *Macri se compara con San Martín* [19], <br> El Destape: *El insólito audio de Faurie a Macri contándole el acuerdo entre Mercosur – Unión Europea* [20], <br> Clarín: *Mauricio Macri, en modo campaña en un acto por el 25 de Mayo: "Volver atrás sería destruirnos"* [21] |
| Elections | Página 12: *Los datos crudos del escrutinio definitivo* [22], <br> La Nación: *Resultados de las Elecciones en Antártida Argentina: seguí el recuento de votos de las PASO* [23], <br> Clarín: *Voto en blanco: no se contará en octubre y hay más chance de definir en primera vuelta* [24] |
| Trash topic | Clarín: *"Esto no es sarasa": el video de Mauricio Macri en la celebración de los 10 años de los centros de Primera Infancia* [25], <br> Clarín: *"Massa: 'Los voy a meter presos'"* [26], <br> Clarín: *La obra en el arroyo el Gato y el "autohomenaje" de Macri* [27] |
| Gossip & Sport | *El deseo de buena suerte de Alessandro Del Piero antes del debut de Daniele De Rossi* [28], <br> *"¿Quién es ese? No lo conozco, lo veré en el Monumental": el Beto Alonso menospreció a Daniele De Rossi* [29], <br> *De Rossi se dio el gusto: cómo vivió su primer partido con Boca en la Bombonera* [30] |

Table 1: Examples of news articles by topic.



\end{document}